\documentstyle[twoside]{article}

\catcode`\@=11
\long\def\@makefntext#1{
\protect\noindent \hbox to 3.2pt {\hskip-.9pt  
$^{{\eightrm\@thefnmark}}$\hfil}#1\hfill}		

\def\@makefnmark{\hbox to 0pt{$^{\@thefnmark}$\hss}}	
	
\def\ps@myheadings{\let\@mkboth\@gobbletwo
\def\@oddhead{\hbox{}
\rightmark\hfil\eightrm\thepage}   
\def\@oddfoot{}\def\@evenhead{\eightrm\thepage\hfil
\leftmark\hbox{}}\def\@evenfoot{}
\def\sectionmark##1{}\def\subsectionmark##1{}}

\oddsidemargin=\evensidemargin
\newcounter{sectionc}\newcounter{subsectionc}\newcounter{subsubsectionc}
\renewcommand{\section}[1] {\vspace{12pt}\addtocounter{sectionc}{1} 
\setcounter{subsectionc}{0}\setcounter{subsubsectionc}{0}\noindent 
	{\tenbf\thesectionc. #1}\par\vspace{5pt}}
\renewcommand{\subsection}[1] {\vspace{12pt}\addtocounter{subsectionc}{1} 
	\setcounter{subsubsectionc}{0}\noindent 
	{\bf\thesectionc.\thesubsectionc. {\kern1pt \bfit #1}}\par\vspace{5pt}}
\renewcommand{\subsubsection}[1] {\vspace{12pt}\addtocounter{subsubsectionc}{1}
	\noindent{\tenrm\thesectionc.\thesubsectionc.\thesubsubsectionc.
	{\kern1pt \tenit #1}}\par\vspace{5pt}}
\newcommand{\nonumsection}[1] {\vspace{12pt}\noindent{\tenbf #1}
	\par\vspace{5pt}}

\newcommand{\textlineskip}{\baselineskip=13pt}
\newcommand{\smalllineskip}{\baselineskip=10pt}

\def\eightcirc{
\begin{picture}(0,0)
\put(4.4,1.8){\circle{6.5}}
\end{picture}}
\def\eightcopyright{\eightcirc\kern2.7pt\hbox{\eightrm c}} 

\newcommand{\copyrightheading}[1]
	{\vspace*{-2.5cm}\smalllineskip{\flushleft
        {\footnotesize Los Alamos archives: gr-qc/9408039 #1}\\
        {\footnotesize $\eightcopyright$\, H.C. Rosu (1998); MPL-A 13, 695-699
        (1998)
         }\\
	 }}

\def\abstracts#1#2#3{{
	\centering{\begin{minipage}{4.5in}\baselineskip=10pt\footnotesize
	\parindent=0pt #1\par 
	\parindent=15pt #2\par
	\parindent=15pt #3
	\end{minipage}}\par}} 

\newcommand{\bibit}{\nineit}

\renewenvironment{thebibliography}[1]
	{\frenchspacing
	 \ninerm\baselineskip=11pt
	 \begin{list}{\arabic{enumi}.}
        {\usecounter{enumi}\setlength{\parsep}{0pt}     
	 \setlength{\leftmargin 12.7pt}{\rightmargin 0pt} 
         \setlength{\itemsep}{0pt} \settowidth
	{\labelwidth}{#1.}\sloppy}}{\end{list}}

\newcounter{itemlistc}
\newcounter{romanlistc}
\newcounter{alphlistc}
\newcounter{arabiclistc}

\def\@citex[#1]#2{\if@filesw\immediate\write\@auxout
	{\string\citation{#2}}\fi
\def\@citea{}\@cite{\@for\@citeb:=#2\do
	{\@citea\def\@citea{,}\@ifundefined
	{b@\@citeb}{{\bf ?}\@warning
	{Citation `\@citeb' on page \thepage \space undefined}}
	{\csname b@\@citeb\endcsname}}}{#1}}

\newif\if@cghi
\def\cite{\@cghitrue\@ifnextchar [{\@tempswatrue
	\@citex}{\@tempswafalse\@citex[]}}
\def\citelow{\@cghifalse\@ifnextchar [{\@tempswatrue
	\@citex}{\@tempswafalse\@citex[]}}
\def\@cite#1#2{{$\null^{#1}$\if@tempswa\typeout
	{IJCGA warning: optional citation argument 
	ignored: `#2'} \fi}}

\def\@refcitex[#1]#2{\if@filesw\immediate\write\@auxout
	{\string\citation{#2}}\fi
\def\@citea{}\@refcite{\@for\@citeb:=#2\do
	{\@citea\def\@citea{, }\@ifundefined
	{b@\@citeb}{{\bf ?}\@warning
	{Citation `\@citeb' on page \thepage \space undefined}}
	\hbox{\csname b@\@citeb\endcsname}}}{#1}}

\def\@refcite#1#2{{#1\if@tempswa\typeout
        {IJCGA warning: optional citation argument
	ignored: `#2'} \fi}}

\def\refcite{\@ifnextchar[{\@tempswatrue
	\@refcitex}{\@tempswafalse\@refcitex[]}}

\def\pmb#1{\setbox0=\hbox{#1}
	\kern-.025em\copy0\kern-\wd0
	\kern.05em\copy0\kern-\wd0
	\kern-.025em\raise.0433em\box0}

\def\fnt#1#2{\footnotetext{\kern-.3em
	{$^{\mbox{\scriptsize #1}}$}{#2}}}

\font\tenrm=cmr10
\font\tenit=cmti10 
\font\tenbf=cmbx10
\font\bfit=cmbxti10 at 10pt
\font\ninerm=cmr9
\font\nineit=cmti9

\font\eightrm=cmr8

\def\qed{\hbox{${\vcenter{\vbox{			
   \hrule height 0.4pt\hbox{\vrule width 0.4pt height 6pt
   \kern5pt\vrule width 0.4pt}\hrule height 0.4pt}}}$}}

\begin{document}



\normalsize\textlineskip
\thispagestyle{empty}
\setcounter{page}{1}

\copyrightheading{}                     

\vspace*{0.88truein}

\centerline{\bf ON THE REMOTE SENSING OF HAWKING GREY PULSES}
\vspace*{0.035truein}
\vspace*{0.37truein}
\centerline{\footnotesize HARET C. ROSU}
\vspace*{0.015truein}
\centerline{\footnotesize\it Instituto de F\'{\i}sica,
Universidad de Guanajuato, Apdo Postal E-143, Le\'on, Gto, Mexico}
\baselineskip=10pt
\vspace*{10pt}
\vspace*{0.225truein}

\vspace*{0.21truein}
\abstracts{
This is a short note on the black hole remote-sensing problem,
i.e., finding out `surface' temperature distributions of various types
of small (micron-sized) black holes from the spectral
measurements of their Hawking grey pulses. Chen's modified
Moebius inverse transform is illustrated in this context.\\
}{}{}


\textlineskip                  
\vspace*{12pt}                 

\vspace*{1pt}\textlineskip	
\vspace*{-0.5pt}
\noindent


\noindent




\noindent





The grey body radiation spectrum from various types of black holes is an
interesting topic, both at the fundamental level$^{1}$
and at the experimental (astrophysical) one.
In a previous paper,$^{2}$ I have applied the modified Moebius transform
(henceforth MMT) due to Chen$^{3}$ to the so-called distorted black holes,
a terminology used by Geroch and Hartle$^{4}$ to denote axially symmetric
black hole Einstein solutions generated by Weyl's technique$^{5}$
from flat space Laplace equation. Here I provide some further details.

In an astrophysical context, one expects
quite naturally to encounter black holes carrying on an ``external"
distribution of matter. Examples are as follows

(i) {\em Weyl black holes}, to which also
Schwarzschild black holes belong. As is well-known
the metric for a static axially symmetric spacetime in Weyl's canonical
coordinates ($t,R,z, \phi$,) reads
$$ds^{2}=-e^{\phi}dt^{2} + e^{\nu - \phi}(dR^{2} +dz^{2}) +
R^{2}e^{-\phi}d\phi ^{2}~,
\eqno(1)
$$
where the metric potentials $\phi$ and $\nu$ are functions of the coordinates
$R$ and $z$ only. The $\phi$ potential is at the same time a solution of the
polar Laplace equation, a remarkable fact discovered by Weyl, while
the $\nu$ potential is easily obtainable from the first one in the form of an
integral.$^{6}$
The astrophysical systems to which these Weyl black holes are mostly
associated, in an already extensive literature,$^{7}$
are giant black hole- disk
(either thin or thick) configurations which are thought to model active
galactic nuclei (AGN). These giant systems will not be included
further in our considerations but one is free to think of such
systems scaled down by many orders of magnitude.

(ii) {\em Black holes with Einstein shells} (again either
thick or thin) which are spherically symmetric, static objects constructed
of test particles travelling along closed geodesics around black holes.
They were first considered by Einstein,$^{8}$ who introduced a well-known
{\em ansatz} for their energy-momentum tensor.

(iii) {\em Primordial black holes} (PBH) and/or any
{\em mini black hole} hovering through the universe and carrying on some
matter distributions. As we already stated, one
may think of the
first two models scaled down by many orders of magnitude,
or of any other conceivable matter distribution accreated around such small
black holes.

(iv) {\em Dirty black holes} in the sense of Visser,$^{9}$ i.e., black
holes in interaction with various classical fields, for which the Hawking
temperature appears to be supressed relative to the vacuum black holes of
equal area.


In the following I shall use the notion of `distorted black holes'
for all these non-isolated black holes. In some cases, the external
distribution of
matter can be of such a kind as to disturb only slightly the pure horizon
Hawking radiation and consequently the character of the problem as an
inverse grey-body radiation problem is preserved. I recall that Hawking
radiation by itself is seriously distorted from a pure black-body radiation,
especially
in the low frequency regime due to a grey-body factor usually identified
with the square of the absorption amplitude for the mode.$^{10}$
Thus, the least our remote sensing arguments require is the common thermal
description of pure black holes, and it is only for astrophysical reasons
that we expect some matter shells (rings, disks, shells, and so on) to occur.

The inverse grey-body problem is a classic one in the field
of remote sensing.$^{11}$
Mathematically, it is usually unstable for most numerical
implementations, i.e., it is an ill-posed inverse problem.$^{12}$ Among the
various procedures to solve it (at least in principle), MMT is a quite
unconventional technique, although it belongs to the Laplace transform
methods.



Referring to the Planck law we shall notice that it actually is
the analytical formula for the power
spectrum, which in laboratory physics is also called
spectral brightness, or spectral radiance of the black body radiation.
The latter notion is used in radiometry to characterize the source
spectral properties as a function of position and direction, which at the
present time are not known experimentally for black holes. For point, i.e.,
far away, grey sources the total radiated power spectrum, also called radiant
spectral intensity is
$$W(\nu)\sim \int _{0}^{\infty} A(T)B(\nu , T)dT~,  \eqno(2)  $$
where $A(T)$ is the area temperature distribution of the grey body, and
$B(\nu , T)$ is the Boltzmann-Planck occupation factor. The inverse grey-body
problem is
to solve the integral equation for $A(T)$ for given total radiated power
spectrum, which may be known either experimentally or within some theoretical
model. This problem was solved in principle by Bojarski$^{13}$
by means of a
thermodynamic coldness parameter $u=h/kT$, and an area coldness distribution
$a(u)$, as more convenient variables to get an inverse Laplace transform of
the total radiated power. The coldness distribution is obtained as an expansion
in this Laplace transform.
More precisely, the total grey power spectrum is rewritten as
$$W(\nu)=\frac{2h\nu ^{3}}{c^{2}}\int _{0}^{\infty}\frac{a(u)}{\exp(u\nu)-1}du
  \eqno(3)  $$
and furthermore as
$$W(\nu) =\frac{2h\nu ^{3}}{c^{2}}\int _{0}^{\infty}\exp(-u\nu)
\sum _{n=1}^{\infty}(1/n)a(u/n)du~.   \eqno(4)$$
Therefore the sum under the integral that we shall denote as $f(u)$ is nothing
else but the Laplace transform of $g(\nu)=\frac{c^{2}}{2h\nu ^{3}}W(\nu)$.

From the Laplace transform $f(u)$ of a grey body, Chen obtained $a(u)$ by
means of the MMT in the form
$$
a(u)=\sum _{n=1}^{\infty}\frac{\mu(n)}{n}f(u/n)~.
\eqno(5)
$$


The Moebius expansion refers to special sums over prime factor-divisors,
(d-sums) of any
number theoretic function $f(n)$, running over all the prime factors of $n$,
1 and $n$ included
$$
f_{1}=\sum_{d|n}^{n} f_{2}(d)~.
\eqno(6)
$$
The remarkable fact in this case is that the last term of the sum can be
written in turn as sums of $f_{1}$ functions, which are called inverse Moebius
transforms (or Moebius d-sums)
$$
f_{2}(n)=\sum _{d|n}^{n} \mu (d)f_{1}(n/d)~,
\eqno(7)
$$
in which $f_{1}(n)$, i.e., the d-sum, becomes the first term of the Moebius
d-sum, and where $\mu(d)$ is the famous Moebius function.
Since each $f_{1}$ term in the second expansion is a d-sum, there is
clearly an overcounting, unless the Moebius functions are sometimes either
naught or negative. The partition of the prime factors of $n$ implied by the
Moebius function is such that, by definition, $\mu(1)$ is 1, $\mu(n)$ is
$(-1)^{r}$ if $n$ includes $r$ distinct prime factors, and $\mu(n)$ is naught
in all the other cases. In particular, all the squares have no contribution to
the inverse Moebius transforms. That is why the integers selected by the
Moebius function are also called square-free integers.

Chen's MMT means to apply such an inversion of finite sums to infinite
summations, and to ordinary functions of real continuous variable(s). MMT
means that if
$$f_{1}(x)= \sum_{n=1}^{\infty} f_{2}(x/n)
\eqno(8)
$$
then
$$f_{2}(x)= \sum_{n=1}^{\infty} \mu (n) f_{1}(x/n)  \eqno(9)  $$
For the inverse grey-body problem, $f_{1}(x)=uf(u)$ and $f_{2}(x)=ua(u)$.
So, one can get the coldness distribution by multiplying the Laplace transform
of the total power spectrum by the coldness parameter, and then applying
the MMT.


I provide now a simple example of the way how the MMT technique works.
For micron-sized Schwarzschild black holes ($M\sim 10^{24}$ g),
no known massive particles are thermally emitted, and according to the
calculations
of Page$^{14}$ about 16\% of the Hawking flux goes into photons, the rest
being neutrino emission. Let us consider these black holes as grey objects,
either by their own,$^{10}$
or even when
possessing some matter distribution close to their horizons. The coldness
parameter will be in the first case
$u_{S}=\frac{1}{\nu}\ln\left(1+
\frac{e^{\beta _{h}\hbar \omega}-1}{\Gamma (\omega)}\right)$, where
$\beta _{h}$ is the horizon inverse temperature parameter,
and $\Gamma (\omega)$ is
the penetration factor of the curvature and angular momentum barrier
around the black hole,
as can be infered from the work of Bekenstein$^{10}$, whereas in the latter
case $u_{S}=h/kT_{d}$, where $T_{d}$ can be considered as an
effective horizon temperature of the distorted black holes
$T_{d}=(8\pi M)^{-1}\exp (2\cal U)$, with
${\cal U}=1/2(1/2\nu - \phi) -\ln(1/2 \sqrt(R/M)$.$^{4}$
Therefore, the coldness distribution will be
$$
a(u_{S})= \frac{c^{2}}{2h\nu ^{3}}
\sum _{n=1}^{\infty} \frac{\mu (n)}{n}f(u_{S}/n)~,
\eqno(10)
$$
where $f$ is the inverse Laplace transform of the total photon power spectrum.

\nonumsection{Acknowledgements}
\noindent
This work was partially supported by the CONACyT Project 4868-E9406.


\nonumsection{References}


\end{document}